\documentclass[10pt,letterpaper,twocolumn,superscriptaddress]{article} 
\usepackage{ol2}
\usepackage[draft]{hyperref}
\usepackage{amsmath}
\usepackage[section]{placeins}
\usepackage{amsmath}
\usepackage{amssymb}
\usepackage{amsfonts}
\newcommand{\op}[1]{\widehat{#1}}                  
\newcommand{\ket}[1]{|#1\rangle}                     
\begin{document}

\twocolumn[ 

\title{Reconstructing the Poynting vector skew angle and wave-front of optical vortex beams via two-channel moir\'{e} deflectometery}

\author{Mohammad Yeganeh$^{1}$, Saifollah Rasouli$^{1,2,*}$, Mohsen Dashti$^{1}$, Sergei Slussarenko$^{3}$, Enrico Santamato$^{3}$, and Ebrahim Karimi$^{4,*}$}

\address{ $^1$Department of Physics, Institute for Advanced Studies in Basic Sciences (IASBS), Gava Zang, Zanjan 45137-66731, Iran \\
$^2$Optics Research Center, Institute for Advanced Studies in Basic Sciences (IASBS),  Gava Zang, Zanjan 45137-66731, Iran \\
$^3$Dipartimento di Scienze Fisiche, Universit\`{a} di Napoli ``Federico II'', Complesso di Monte S. Angelo, 80126 Napoli, Italy\\
 $^4$Department of Physics, University of Ottawa, 150 Louis Pasteur, Ottawa, Ontario, K1N 6N5 Canada\\
$^*$Corresponding authors: rasouli@iasbs.ac.ir and ekarimi@uottawa.ca
}

\begin{abstract}
A novel approach based on the two-channel moir\'{e} deflectometry has been used to measure both wave-front and transverse component of the Poynting vector of an optical vortex beam. Generated vortex beam by the $q$-plate, an inhomogeneous liquid crystal cell, has been analyzed with such technique. The measured topological charge of generated beams are in an excellent agreement with theoretical prediction.
\end{abstract}

\ocis{050.4865, 010.7350, 120.4120, 260.6042.}

 ] 

\noindent It is well-understood that optical beams may possess two \textit{rotational} like degrees of freedom corresponding to vector and phase features of the light that are known as a spin and orbital angular momentum~\cite{allen:03}. A beam with circular polarization possesses spin angular momentum (SAM) of $\pm \hbar$ per photon, which sign is defined by the polarization handedness. In contrast, orbital angular momentum (OAM) is associated with helical phase-front of the beam. The optical field is proportional to $\exp{(i\ell \phi)}$ and carries $\ell\hbar$ angular momentum per photon -- $\phi$ and $\ell$ stand for azimuthal angle and integer number (winding number), respectively~\cite{frankearnold:08}. Laguerre--Gaussian modes are an example of paraxial optical beams with well-defined value of OAM.  For non-zero $\ell$, the beam has no light at the center and the intensity profile has a doughnut shape~\cite{siegman}. In a full rotation around such dark point (\textit{singular point}) the optical phase suffers a jump of $\pm 2|\ell|\pi$, where the sign determines the phase rotation direction. In paraxial beams the SAM and OAM correspond to different physical properties and are independent rotational degrees of freedom~\cite{allen:03}. In recent times, such features paved the way to promising applications both in the classical and quantum optics such as: imaging techniques, lithography, optical tweezers, and communications~\cite{he:95,gibson:04,molina:07,nagali:09np}. A well known feature of light beams carrying OAM is that the Poynting vector owns non-negligible transverse components. The radial and azimuthal components of the Poynting vector  density (Poynting vector/beam intensity) in the beam transverse plane tend to zero and infinity, respectively, as the central dark singularity is approached. In particular, a collimated helical beam bears an azimuthal component of the Poynting vector proportional to the winding number $\ell$~\cite{miles:95}.  Different approaches have been already implemented to measure, directly or indirectly, the transverse components of the Poynting vector or the wavefront profile of a helical beam~\cite{simpson:97}. For example, a Shack-Hartmann sensor was used to measure the skew angle of the Poynting vector and exploited as an OAM sorter~\cite{leach:06,starikov:07}. However, local precision, low transverse resolution and aberrations bound its utilization to few pedagogical and research applications. \newline
In this letter, we propose a novel approach based on two-channel moir\'{e} deflectometer to measure both phase-front and transverse components of the Poynting vector of an optical vortex (OV) beam~\cite{rasouli:10ol}. Unlike the interferometric approach, the moir\'{e} technique does not require a plane wave reference beam. Furthermore, the accuracy and measurement quality compared to the wave-front based sensors, e.g. Shack-Hartmann, have been improved significantly.\newline

Among OV beams, LG modes play an important role. They are a solution to the paraxial wave equation (PWE) that form orthogonal, complete set, and are eigenstates of the light OAM. Any paraxial optical beam can be expanded in the LG basis~\cite{siegman} so that exploiting the physical features of LG modes provides a profitable knowledge about general properties of OV beams. Thereafter, with little loss of generality, a pure LG beam was considered. A straightforward calculation shows that for such a beam, the wavefront and  Poynting vector skew angle at the waist are given by $\chi=\ell\phi$ and $\alpha=\ell/({k \rho})$, respectively - where $\rho, \phi, z$ are the cylindrical coordinates, $k=2\pi/\lambda$ is the wavenumber, and $\alpha$ stands for the ratio of azimuthal and longitudinal component of Poynting vector~\cite{miles:95}. The slanting angle  $\alpha$ increases for small radii, which is an evidence of staircase wavefront feature $\mathbf{k}\propto \mathbf{\nabla}\chi $. Of course, it is worth to remember that, due to the presence of phase singularity, the optical phase is undefined at the origin and the light intensity drops to zero as well. \newline
In our work, we generate OV beams by means of $q$-plate, a liquid crystal cell possessing a specific integer or half-integer positive or negative topological charge $q$ at its center~\cite{marrucci:06prl}. When the optical retardation of the $q$-plate is $\pi$, \textit{tuned $q$-plate}, the handedness of the input circular polarization is reversed and the OAM value of the input beam is increased/decreased by $2q\hbar$ per photon:
\begin{eqnarray}\label{eq:U}
	\op{\hbox{U}}_{\hbox{$q$-plate}}\cdot\ket{\pm}=e^{\pm 2iq\phi}\ket{\mp},
\end{eqnarray}
where $\ket{\pm}$ stand for left and right circular polarization and $q$ is the $q$-plate topological charge (unessential global phase factor has been omitted)~\cite{marrucci:11jo}. In a tuned $q$-plate, the handedness of the OAM added to the input beam depends on its helicity as shown in Eq.~(\ref{eq:U}). The optical retardation of the $q$-plate can be practicality adjusted by varying temperature, by applying pressure or, more conveniently, by applying an AC voltage~\cite{karimi:09apl,slussarenko:11}.  It is worth noting that $q$-plates as other available OAM generators do not generate pure LG beam~\cite{karimi:09}. Indeed, they generate Hypergeometric-Gaussian beam, which can be seen as a superposition of infinite LG$_{p,\ell}$ modes, all sharing the same winding number $\ell$ and differing in the radial index $p$~\cite{karimi:07}. However, it can been shown that the most of the beam power is confined in the LG$_{0,\ell}$ mode.\newline

In our experiment, a collimated plane wave of a He-Ne laser ($\lambda=633$~nm, $\rm{P=30~mW}$) was converted to an OV beam with OAM of $\ell=\pm20$ by means of an electrically controlled $q$-plate with $q=+10$~\cite{slussarenko:11}. The OV beam was then split out into two beams, going up and down, say, by a beam splitter (BS). A mirror was used to introduce even number of reflections in the reflected beam so that the light beams in both arms had equal OAM value. Each beam in the two arms passed through a pair of \textit{moir\'{e} deflectometers}, which were placed parallel and close to each other. The optical path lengths of the two arms were taken equal to obtain uniform mapping scale in the transverse plane. Directions of the grating's rulings were almost parallel in each moir\'{e} deflectometer -- but, the gratings of the deflectometers in the two arms were perpendicular each other (see Ref.~\cite{rasouli:10oe} for more detail on the experimental setup).  We assume the grating rulings in the first and second channels are along the $x$- and $y$-directions, respectively.  Distances in each channel were set to have the Talbot image of the first grating formed at the second grating position, where a diffuser was installed. The second gratings were slightly rotated to obtain moir\'{e} effect.  The gratings had equal periods of $d=0.1~mm$ and the used Talbot distance was set at $Z_{k}=6.34~cm$, corresponding to $2^{nd}$ order. The moir\'{e} patterns from both arms, which consisted of 317$\times$352 pixels each, were finally acquired by CCD camera for further processing~\cite{rasouli:10oe}. The sensitivity of our apparatus allowed us to measure phase-front angle changes of $7.77\mu$rad.
\begin{figure}[t]
\begin{center}
	\includegraphics[width=8.cm]{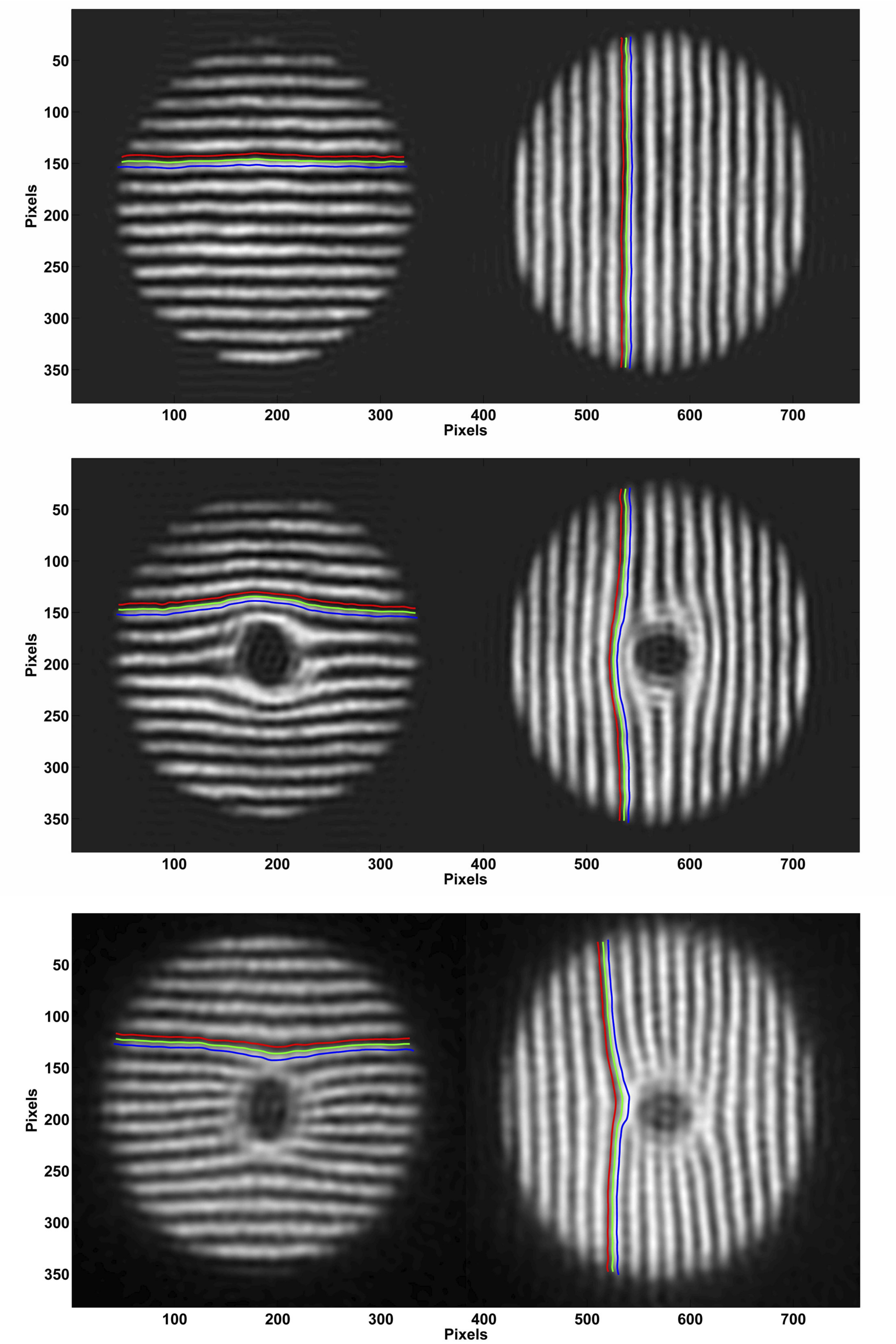}
	\caption{\label{fig:moire-ccd} (Color online) A typical recorded frame for three different cases (upper row) plane wave, (middle row) OV beam with OAM of $+10$ and (down row) OV beam with OAM value of $-10$, respectively. Each frame consists of two set of orthogonal moir\'e patterns corresponding to the beams in two arms. Red,  green, and blue solid lines stand for typical traces of moir\'{e} fringe minima, fringe maxima, and first order virtual trace, respectively.}
\end{center}
\end{figure}
Fig (\ref{fig:moire-ccd}) shows the recorded frames of two moir\'{e} patterns on the CCD camera for three different cases: \textbf{(i)} plane wave (when the $q$-plate is totally untuned -- $2\pi$ optical retardation -- the emerged TEM$_{00}$ beam was selected by an appropriate polarization selector), \textbf{(ii)} OV beams having OAM $\ell=+10$ and \textbf{(iii)} $\ell=-10$ generated by a tuned $q$-plate, respectively. Each frame consists of two moir\'e fringe patterns along $x$ and $y$-directions corresponding to the beams in the two arms of the apparatus~\cite{rasouli:10oe}. In the first case, the wavefront of the plane TEM$_{00}$ beam, was planar. Therefore, the moir\'e fringe patterns on both arms were uniformly spaced linear dark and bright fringes. In the last two cases, where beams possess helical phase front, displacements in the moir\'{e} fringes have been observed in both channels. Such fringe deviations reflect the beam wave-front curvature and we exploited such changes to determine the shape of the incoming beam phase-front and the transverse component of the Poynting vector. A straightforward calculation shows that the wavefront gradient $\nabla\chi$ can be expressed in term of the relative fringe displacements in the two arms of our apparatus:
\begin{equation}
	\mathbf{\nabla\chi}=\left(\frac{d}{d_m^{'}Z_k} \right)\delta y_m\,\hat{x}+\left(\frac{d}{d_m Z_k}\right)\delta x_m\,\hat{y},
\end{equation}
where $d_m$, $d'_m$ are the moir\'{e} fringe spacings, and $\delta x_m$, $\delta y_m$ are the moir\'{e} fringe shifts in the first ($x$-direction) and second ($y$-direction) channels, respectively~\cite{rasouli:12}.
\begin{figure}[h]
\begin{center}
	\includegraphics[width=8.5cm]{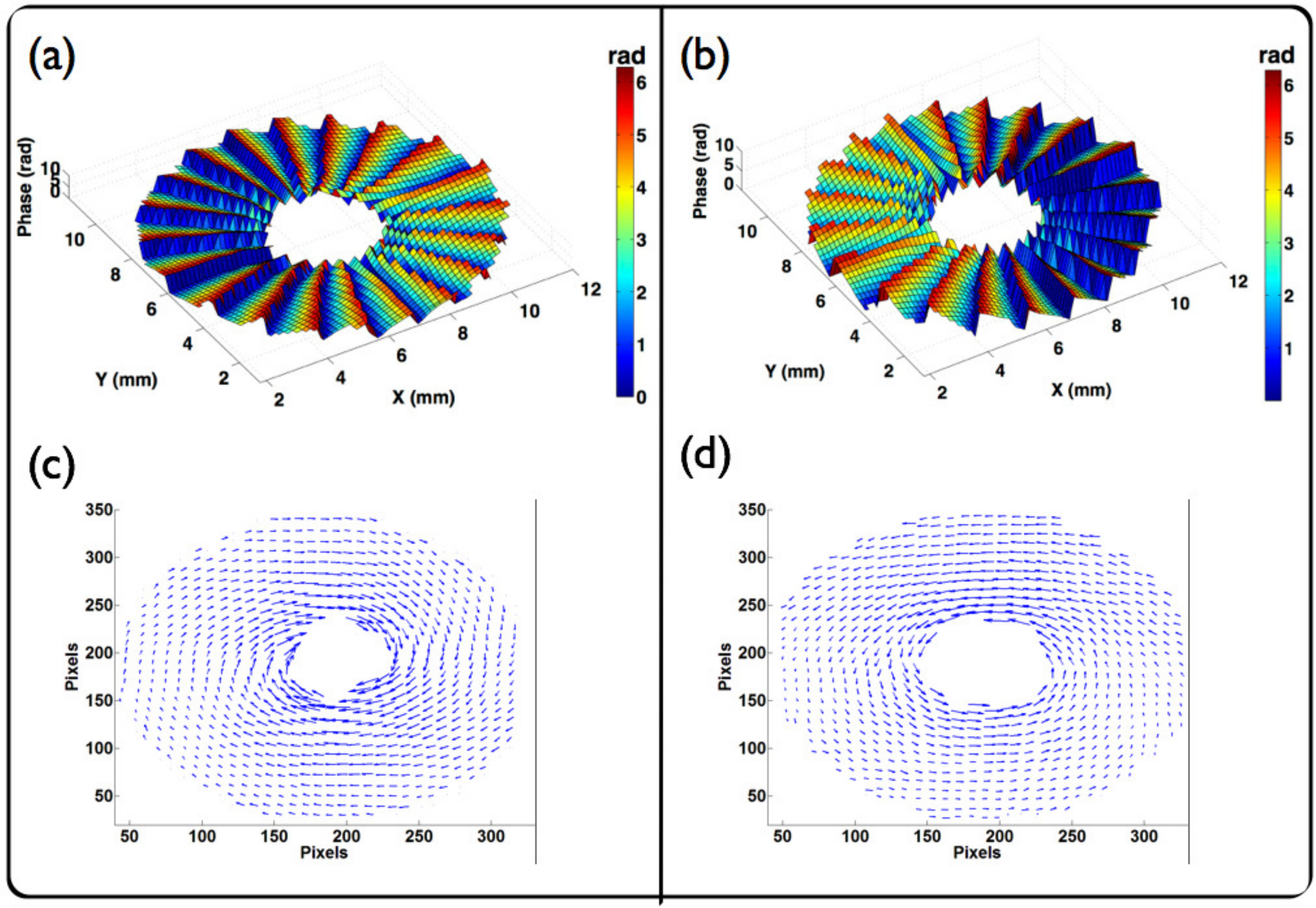}
	\caption{\label{fig:wf_pv} (Color online) The reconstructed wave-front and transverse component of the Poynting vector of OV beams. (a) and (c) are the reconstructed wave-front and transverse component of Poynting vector for an OV beam with $\ell=+20$, respectively and (b) and (d) are the same for an OV beam with $\ell=-20$. The beam propagation direction is toward the plane, and the phase value scale is shown in false color in the side bar.}
\end{center}
\end{figure}
The reconstructed beam wave-front and transverse Poynting vector of OV beams with topological charge $\pm20$ are shown in Fig. \ref{fig:wf_pv}. The central region below a certain threshold was excluded due to the presence of the singularity. As shown in the figure, the transverse components of the Poynting vector circulates around the singular point and its value increases at inner radii as expected from theory. In order to experimentally prove the presence of vortex structure in the light beam, the beam topological charge ${\cal Q}=\frac{1}{2\pi}\oint_C \nabla\chi\cdot d\mathbf{r}$ with loop $C$ enclosing the singularity, was calculated from the data~\cite{allen:03}. We found a topological charge of $19.93\pm 0.08$ in excellent agreement with the predicted value~\cite{marrucci:11jo}. The reported error corresponds to different choices of the contour loop $C$. In a different approach, the dependence of the azimuthal component $S_\phi$ of the Poynting vector on $\ell$ ($S_\phi\propto\ell/\rho$~\cite{miles:95}) can be used to measure the beam OAM. Therefore, a direct plot of the skew angle $\alpha=\ell/(k\rho)$ of Poynting's vector for different $\rho$ unveils the beam $\ell$.\newline

In conclusion, we presented a novel approach based on the moir\'{e} deflectometry to measure both wave-front and transverse component of Poynting vector in OV beams. Although not interferometric, such method leads to a very precise direct measurement of skew angles of Poynting's vector with accuracy of $7.77\mu$rad. This accuracy can be improved by exploiting a lower grating period (i.e. $d=1/50$~mm). In our experiment, we have reconstructed the beam wavefront and the skew angle of Poynting's vector for two different OV beams generated by $q$-plate. We think that the present technique could be very promising to measure the OAM, topological charge of light vortex beam and components of the Poynting vector~\cite{angelsky:12,angelsky:11}, since it is not interferometric, insensitive to the precise location of the optical vortex in the transverse plane and much more accurate than other current available approaches.\newline

M.Y., S. R. and M. D. acknowledge the IASBS research council for partial financial support, S. S. and E. S. acknowledge the financial support of the Future and Emerging Technologies (FET) programme within the Seventh Framework Programme for Research of the European Commission, under FET-Open grant number 255914-PHORBITECH, and E. K. acknowledges the support of the Canada Excellence Research Chairs (CERC) program.


\end{document}